\documentclass[10pt,twocolumn]{article}
\usepackage{dsfont}      
\usepackage{boxedminipage}
\usepackage{amsfonts}
\usepackage{amssymb}
\usepackage{amsmath}     
\usepackage{epsfig,changebar,eepic,psfig}
\usepackage{graphicx}

\usepackage{tmcomplexityclassessw}
\usepackage{specialletterssw}
\usepackage{capitalosw}
\usepackage{algebrasw}
\usepackage{functionssw}
\usepackage{deviceoperatorssw}
\usepackage{bddsw}
\usepackage{theoremssw}

\usepackage{colordvi} 

\usepackage[dvips,dvipsnames]{color} 

\newcommand{\llabel}[1]{\label{#1}}

\newcommand{\inputs}{\ensuremath{\{ 0,1 \}}}

\newcommand{\comment}[1]{}

\newcommand{\ket}[1]{|#1\rangle}
\newcommand{\bra}[1]{\langle #1\,|}
\newcommand{\braket}[2]{\langle #1\,|\,#2\rangle}

\newcommand{\nosubset}{\text{ is a proper subset of }}

\DeclareMathOperator{\fa}{finalAmp}

\begin{document}
\bibliographystyle{plain}

\title{Quantum Ordered Binary Decision Diagrams\\ with Repeated Tests}            
\author{Matthias Homeister\and Stephan Waack}
\date{\small Institut f\"ur Numerische und Angewandte Mathematik\\
Georg-August-Universit{\"a}t G\"ottingen\\
Lotzestr.~16--18, 37083 G{\"o}ttingen, Germany\\
  \texttt{\{homeiste,waack\}@math.uni-goettingen.de}}
\maketitle{}
\thispagestyle{empty}
\begin{abstract}
Quantum branching programs (quantum binary decision diagrams, respectively) are a convenient tool for examining 
quantum computations using only a logarithmic amount of space. 
Recently several types of restricted quantum branching programs
have been considered, e. g. read--once quantum branching programs.
This paper considers 
quantum ordered binary decision diagrams (QOBDDs) and answers the
question: 
How does the computational power of QOBDDs increase, if we allow
repeated tests. 
Additionally it is described how to synthesize QOBDDs according to Boolean operations.

\end{abstract}

\section{Introduction}\label{introductionSection}

A central question of quantum complexity is, in 
which cases quantum computations do outperform classical ones. 
Famous results are the algorithms of Shor (\cite{shor97}) and Grover
(\cite{gro96}); 
apart from that much has been achieved by examining various restricted
models of quantum computation 
and comparing them with their classical counterparts. One such model are  
\emph{branching programs} (BPs). 
They are related to circuits, Boolean formulas, and nonuniform
space complexity. Quantum branching programs have been considered in 
\cite{Naka00}, \cite{Abla01}, \cite{Spa02}, \cite{AMP02}, \cite{Abla03} and \cite{Sau04}.

A \emph{deterministic} BP $B$ 
on the variable set $\{x_1,x_2,\ldots, x_n\}$ consists of a
directed acyclic graph $G=(V,E)$ where multi-edges are allowed.
Two of the nodes are denoted as \emph{targets}.
They are sinks in the graph-theoretical sense, and are labeled $0$ and $1$.
The other nodes are called \emph{branching nodes}.
They get labels from 
$\{x_1,x_2,\ldots, x_n\}$.
The edges get labels from $\{0,1\}$. 
For each branching node, there is exactly one outgoing edge labeled $0$, and
one outgoing edge labeled $1$.
The \emph{size} of a BP is the number of its edges.

A branching program computes a Boolean function in a natural way. 
Each input $a\in\{0,1\}^n$ \emph{activates} all $a_i$-edges leaving
$x_i$-nodes, for $i=1,2,\ldots n$.
A \emph{path} in $G$ is defined to be activated by $a$, if $a$ activates all
its edges. 
An input $a$ is \emph{accepted} if the path activated by $a$ leads to
the $1$-sink 
and is \emph{rejected} in the other case.
This model can be generalized to nondeterministic or probabilistic
modes of computation in a straightforward way, 
see \cite{Weg2000}. 

\emph{Quantum branching programs} (QBPs)
 can be defined by adding transition amplitudes to the edges and
 allowing more than two sinks (see \cite{Sau04}). 
We outline this approach very briefly. The computation on an input $a$
 starts at the source of the QBP. 
With respect to the transition amplitudes each step of the computation consists of a superposition of nodes.
Finally a measurement determines the result of the computation,
i.e. the label of the resulting sink. 
Certainly the transition amplitudes have to fulfill a 
global well--formedness constraint that ensures a unitary evolution of the computation.

In this paper we consider another -- equivalent -- approach following
\cite{Abla01}. 
This approach is particularly useful for \emph{leveled} branching
programs, 
where the nodes are partioned into levels $L_1,\ldots,L_m$. 
$L_m$ consists of the sinks, for $i\le m$, the nodes in $L_i$ are
labeled by the same variable and 
the outgoing edges of a node in $L_i$ lead to nodes in the level below, i.e. in $L_{i+1}$.

A quantum branching program on the variables $\{x_1,x_2,\ldots, x_n\}$ of width $w$ and length $l$ consists of 
\begin{itemize}
\item a set $D$ of cardinality $w$; we assume $D=\{\ket{1},\ldots,\ket{w}\}$,
\item a sequence of pairs of unitary tranformations
  $(T^0_{y_i},T^1_{y_i}), i=1,\ldots,l$ 
on the complex vector space spanned by $D$
 where $y_i$ is a variable in $\{x_1,x_2,\ldots, x_n\}$
\item a starting state $\ket{1}\in D$ and 
\item a set of accepting states $F\subseteq D$.
\end{itemize}

We call the sequence $(y_1,\ldots,y_l)$ of variables in
$\{x_1,x_2,\ldots, x_n\}$ 
the \emph{variable ordering} of the QBP. 
The computation proceeds in the complex vector space spanned by $D$. 
Each state of the computation is a vector of length $1$ in this space.
 The computation starts with state $\ket{1}.$ On an input
 $a\in\inputs^n$ the transformation $T^0_{y_1}$ 
is applied if $a$ assigns  the variable $y_1$ to $0$, 
in the other case $T^1_{y_1}$ gets used. 
The result is a unit vector on that either $T^0_{y_2}$ or $T^1_{y_2}$
is applied. 
After $l$ steps the computation stops by \emph{measuring} the state 
$$
\ket{\psi_{l}(a)}=T^{\epsilon_l}_{y_l}\cdot\ldots \cdot
T^{\epsilon_2}_{y_2}\cdot T^{\epsilon_1}_{y_1}\ket{1}, 
$$
where $a$ assigns  $y_i$ with  $\epsilon_i$, and  
$\ket{\psi_{l}(a)}$ is a vector $(\alpha_1,\ldots,\alpha_w)$ whose
components are complex amplitudes, 
or -- equivalently -- the superposition $\sum_{i=1}^w\alpha_i\ket{i}$. 
The measurement results with probability $|\alpha_i|^2$ in the state
$\ket{i}$. 
If this result is a member of $F$ we accept the input, in the other case we reject.  
Therefore the state $\ket{\psi_{l}(a)}$ plays an important role --
this leads to the definition of the 
\emph{final amplitude} in a computation. 
Let \CB\, be the QOBDD defined above. 
Then the final amplitude of  $\ket{i}$ according to the input $a$ is 
$$\fa(\ket{i},a):=\braket{i}{\psi_{l}(a)},$$
 i.e. the component of $\ket{i}$ just before the measurement; 
$\braket{i}{j}$ denotes the inner product of the complex vectors
 $\ket{i}\ket{j}$. 
For each $\ket{i}\in D$ the measurement finishing the computation of
 \CB\, 
on $a$ yields the result $\ket{i}$ with probability
 $|\fa(\ket{i},a)|^2$. 
An input $a$ is accepted with probability $$\sum_{i\in F}|\fa(\ket{i},a)|^2.$$ 
Sauerhoff and Sieling proved in \cite{Sau04} that QBPs of polynomial
 size correspond to 
logarithmic space restricted computations of nonuniform quantum Turing
 machines. In our model the \emph{size} is the product of width and
 length. 
Ablayev, Moore and Pollett proved that NC$^1$ can be accepted 
by QBPs of width $2$ and polynomial length, see \cite{AMP02}. 
Upper and lower bounds have been proved for several restricted versions. 
An important variant are \emph{quantum ordered binary decision diagrams} (QOBDDs). 

Branching programs are important not only in theory but also in applications.
In this context they are denoted as
binary decision diagrams (BDDs).
BDD-based data structures for Boolean functions play a key role in
hardware verification, test pattern generation, symbolic simulation,
logical synthesis or analysis, and design  of circuits and automata
(for a survey see \cite{Weg2000}).
Once the model is chosen one needs efficient algorithms for many
operations, particularly for synthesis, minimization
and equivalence test.
The non-equivalence test for two functions $f$ and $g$ is equivalent
to the satisfiability problem for $f\oplus g$.
It is known that the satisfiability problem for read-twice branching programs
is \np-complete.
Therefore one prefers the restricted types of branching
programs.
Very important is one introduced by Bryant \cite{Bry86} 
that may be regarded as the state-of-the-art data structure in many
applications.

A QBP as defined above is a \emph{QOBDD} if the 
variable ordering $(y_1,\ldots,y_l)$ is a permutation of
$\{x_1,x_2,\ldots, x_n\}$. 
Or, more illustrative, the length is $n$ and each uniform transformation depends on another variable.

QOBDDs have been considered by
Sauerhoff and Sieling in \cite{Sau04}. 
They have presented a function computable by succinct QOBDDs that
requires exponential size deterministic OBDDs. 
Counterwise they have found a very simple function that is not computable by polynomial size QOBDDs. 
They call this function NO$_n$ (neighbored ones). 
It is defined on $n$ variables $\{x_1,x_2,\ldots, x_n\}$ 
and tests whether there are neighbored variables with value $1$, 
i.e. an input is accepted if and only if $x_i=x_{i+1}=1$ for some $i<n$.
 This function is computable by deterministic OBDDs of size $\Oh{n}$. 
This weakness of QOBDDs has the reason that every step of the
computation is a unitary and therefore reversible transition. 
For strongly restricted models of quantum computations 
(the situation is similar for some kinds of quantum finite automata) 
it seems to be difficult to forget variables already read (see \cite{Sau04}).

Thus the question arises: How does this situation change 
if we slightly diminish the restriction? 
How does the computational power change if 
we consider QOBDDs with repeated tests? This leads to the definition of \emph{$k$-QOBDDs}.

Unformally a $k$-QOBDD is the concatenation of $k$ QOBDDs that obey the same variable ordering. 
More presicely a $k$-QOBDD \CB\, according to the variable ordering
$\sigma$ 
is a QBP \CB\, on the variables $\{x_1,\ldots,x_n\}$ of length $kn$.
The levels are partitioned into $k$ layers of length $n$. 
In each layer the variables are tested according to $\sigma$. 
The computation of a $k$-OBDD on an input $a$ 
is determined by a sequence of unitary transformations 
$T^\epsilon_1,\ldots,T^\epsilon_{kn}, \epsilon\in\{0,1\}$, 
where $T^\epsilon_i$ is chosen according to the input bit $a_{\sigma(i)}$. 
Thus, on input $a$ the $i$-th layer performs the computation 
$T^{a_{\sigma(n)}}_{i\cdot n}\cdot \ldots\cdot
T^{a_{\sigma(2)}}_{(i-1)\cdot n+2}
\cdot T^{a_{\sigma(1)}}_{(i-1)\cdot n+1}$.

Bollig, Sauerhoff, Sieling and Wegener have proved in \cite{BSSW98} 
a hierarchy on deterministic $k$-OBDDs. 
It turns out that the computational power of polynomial size $k$-OBDDs is strictly greater than that of $(k-1)$-OBDDs.
Things are different in the case of nondeterministic $k$-\OBDD s.
In \cite{BHW04} it is shown that nondeterministic, 
parity and randomized $k$-OBDDs are not more powerful than OBDDs with the correspondent computation modes.
 
Section \ref{sec2} starts 
with an analysis of the way a $k$-QOBDD computation on an input $a$
evolves. 
To apply this analysis for comparing QOBDDs with $k$-QOBDDs, 
we consider products of QOBDDs in \ref{synsec}. 
This provides a method to perform the Boolean synthesis. 
Section \ref{ppsec} shows that repeated tests are of no use in the case of QOBDDs with unbounded error.

\section{The way a $k$-QOBDD computes}\llabel{sec2}

We consider a $k$-QOBDD \CB\, with variable 
ordering $\sigma$ and width $w$ on $n$ variables. 
The computation evolves according to the unitary transformations 
$T^\epsilon_1,\ldots,T^\epsilon_{kn}, \epsilon\in\{0,1\}$. 
We define $U_i(a)$ to be the transformation performed by the $i$-th
layer under $a$. 
Formally, 
$U_i(a)=T^{a_{\sigma(n)}}_{i\cdot n}\cdot \ldots\cdot
T^{a_{\sigma(2)}}_{(i-1)\cdot n+2}
\cdot T^{a_{\sigma(1)}}_{(i-1)\cdot n+1}.$
The final amplitudes of the computation on $a$ are the components 
of the superposition $U_k(a)\cdot\ldots\cdot U_1(a)\ket{1}$. 

Let $\alpha^{(\lambda)}_{i j}(a)$ be the amplitude of $\ket{j}$ 
in the state $U_\lambda(a) \ket{i}$, i.e. $$\alpha^{(\lambda)}_{i j}(a)=\bra{j}U_\lambda(a) \ket{i}.$$
We define the column vector $\mu^{(1)}$ 
of length $w$ of functions from $\{0,1\}^n$ to $\IC$ by
\begin{equation}
\mu^{(1)}_j(a):=\alpha^{(1)}_{1 j}(a),\qquad j\in\{1,2,\ldots, w\}\label{mu1DefEQ}.
\end{equation}

For $\lambda\in\{2,3,\ldots, k\}$, let $\mu^{(\lambda)}$ be a 
  $w\times w$-matrix  of functions from $\{0,1\}^n$ to $\IC$\, defined by
\begin{equation}
   \mu^{(\lambda)}_{i,j}(a):=\alpha^{(\lambda)}_{i j}(a),
       \quad i,j\in\{1,2,\ldots, w\} \label{muDefEQ}.     
  \end{equation}

According to our definition, $\mu^{(\lambda)}_{i,j}(a)$ equals 
the amplitude of $\ket{j}$ in the result of the computation of the $\lambda$-th layer on $a$ starting with $\ket{i}$.
 
For $\ket{i}\in D$ we define $\beta_i:=\fa(a,\ket{i})$.   
An easy calculation reveals that for all $a\in\{0,1\}^n$
the vector of final amplitudes $\beta_i$ can representated 
as a matrix product:
\begin{multline}\llabel{accMatrixRepEQ} 
 (\beta_1(a),\ldots,\beta_w(a))^T = \\
\mu^{(1)}(a)^T \cdot \mu^{(2)}(a)\cdot\ldots
\cdot\mu^{(k-1)}(a) 
\cdot\mu^{(k)}(a)
\end{multline}

Figuring out the right hand side of Equation
\ref{accMatrixRepEQ}, we obtain for $j=1,\ldots,w$
\begin{multline}
 \beta_j(a) = \\
  \sum_{i_2,\ldots,i_k\in \{1,\ldots,w\}}
  \underbrace{
    \mu^{(1)}_{i_2}(a)\cdot \mu^{(2)}_{i_2i_3}(a)\cdot\ldots\cdot
    \mu^{(k)}_{i_kj }(a) 
    }_{=:\mu_{i_2i_3\ldots i_k j}(a)}.\llabel{accProbSumRepEQ} 
\end{multline}

We define the acceptance probability of \CB\,  on some input $a$ by $$acc(a):=\sum_{\ket{j}\in F} |\beta_{j}|^2.$$
Our purpose is to construct a quantum OBDD \CB'\, 
that simulates the quantum $k$-OBDD \CB\, 
in the case of bounded error computations. \CB'\, 
accepts an input $a$ with probability greater than $1/2$ if and only if $acc(a)>1/2$.
To this end we adopt the well-known ``product-construction''
for finite automata common for synthesizing BPs. 

\section{Product construction and Boolean synthesis}\llabel{synsec}

In the quantum case it is convenient to use the tensor product. 
Let $\CB_1$ and $\CB_2$ be quantum OBDDs using the transformations 
$(T^0_{x_i},T^1_{x_i}), i=1,\ldots,n$ and $(S^0_{x_i},S^1_{x_i}),
i=1,\ldots,n$, 
respectively (the same approach works for $k$-QOBDDs).
$\CB_i$ is defined on the set $D_i$ of cardinality $w_i$, $i=1,2$.
We denote 
$$\CB_1\otimes\CB_2$$
 to compute on the set $D_1\times D_2$ of elements $\ket{i}\otimes\ket{j}$ by the transformations 
$$(T^0_{x_i}\otimes S^0_{x_i},T^1_{x_i}\otimes S^1_{x_i}),\qquad i=1,\ldots,n.$$ 
Let $\ket{d_i}$ be contained in $D_i, i=1,2$. 
The complex values $\fa(\ket{d_i},a)$ are the according final amplitudes of the computations of $\CB_i, i=1,2$.

 Then for the computation of $\CB_1\otimes\CB_2$ it holds that 
\begin{multline}\llabel{prodEq}
\fa(\ket{d_1}\otimes \ket{d_2},a)=\\
\fa(\ket{d_1},a)\cdot \fa(\ket{d_2},a).
\end{multline}

Using Equation \ref{prodEq} and standard techniques as
OBDD-probability-amplification 
we can prove that the logic synthesis operation is feasible for QOBDDs with different error bounds.

\begin{proposition}\llabel{synProp}
Let $f_i, i=1,2$ be functions computable by quantum OBDDs of width
$w_i$. 
Then $f_1\wedge f_2$ is computable by a quantum OBDD of width polynomial in $w_1w_2$.
\end{proposition} 

The proof of Proposition \ref{synProp} is straightforward. 
We define the accepting states $F^\otimes$ as $F_1\otimes F_2$. 
If $\CB_i$ accepts an input $a$ with probability $p_i, i=1,2$ then
$\CB_1\otimes\CB_2$ 
accepts $a$ with probability $p_1p_2$. 
Thus synthesizing two QOBDDs with error bound $\epsilon$ result in a 
QOBDD computing the conjunction of the input QOBDDs with error bound
$1-(1-\epsilon)^2$. 
By a finite number of additional synthesis steps the error can be decreased to $\epsilon$. 

\section{Quantum $k$-OBDDs with unbounded error}\llabel{ppsec}   

We make use of the product construction described in the 
preceding subsection to simulate a $k$-QOBDD by a QOBDD. 
Consider the $k$-wise product of $D=\{\ket{1},\ldots,\ket{w}\}$, i.e.
$$D^\otimes:=\bigotimes_{i=1}^k D=\{\ket{i_1i_2\ldots i_k};\ket{i_l}\in D, l=1,\ldots,k\}.$$
$\ket{i_1i_2\ldots i_k}$ is the common abbreviation of $\ket{i_1}\otimes\ldots\otimes\ket{i_k}$.
To simulate the $k$-QOBDD \CB\, we define a QOBDD $\CB^\otimes$\, 
computing on the set $D^\otimes$. Its transformations are 
$(T_1^{\otimes,0},T_1^{\otimes,1}),\ldots,(T_n^{\otimes,0},T_n^{\otimes,1})$ 
where $T_i^{\otimes,\epsilon},\epsilon\in\{0,1\}$ are chosen according to $x_{\sigma(n)}$.
We define 
\begin{equation}
T_i^{\otimes,\epsilon}=T_i^{\epsilon}\otimes
T_{i+n}^{\epsilon}\otimes\ldots\otimes T_{i+(k-1)n}^{\epsilon},\llabel{tcrossEq}
\end{equation}
for $\epsilon\in\{0,1\}, i=1,\ldots,n.$ 
Note that on some input $a$ the QOBDD $\CB^\otimes$ 
performs the unitary transformation $U^\otimes(a)=U_1(a)\otimes\ldots\otimes U_k(a)$.

Let us examine how $\CB^\otimes$\, simulates the way \CB\, computes. Let $a\in\inputs^n$ be fixed.
We apply $U^\otimes(a)$ on $\ket{1i_2i_3\ldots i_k},$ where $i_2i_3\ldots i_k$ are arbitrarily chosen elements of $D$. Let
\begin{multline}\llabel{psitimes}
\psi_{1i_2i_3\ldots i_k}(a):=\\
U^\otimes(a)\ket{1i_2i_3\ldots i_k} =\\
U_1(a)\ket{1}\otimes U_2(a)\ket{i_2}\otimes\ldots \otimes U_k(a)\ket{i_k}
\end{multline}

We start the computation in state $\ket{1i_2i_3\ldots i_k}$ as above.
then the component $\ket{i_2i_3\ldots i_kj}$ for $\ket{j}\in D$ of the 
state $\psi_{1i_2i_3\ldots i_k}(a)$ has the same amplitude as thefollowing 
computation path $\pi$ of \CB. $\pi$ starts with $\ket{1}$, 
the intermediate result after the first layer is $i_2$, after the second layer we reach $i_3$ etc; 
the final result of the considered path $\pi$ is $\ket{j}$. 
Note, that it is quite natural to carry the concept of a 
superposition of states to a superposition of computation paths. 
Formally it holds that
\begin{equation}
\braket{i_2i_3\ldots i_kj}{\psi_{1i_2i_3\ldots i_k}(a)}=\mu_{i_2i_3\ldots i_k j}(a),\llabel{psiEq}
\end{equation}
using the notation of Equation \ref{accProbSumRepEQ}.

We utilise Equation \ref{psiEq} to build a QOBDD \CB' that performs the same computation as the $k$-QOBDD \CB.
We define $D':=D^\otimes\cup\{\ket{t_0},\ket{t_1}\}$. 
$t_1$ is an accepting state and $t_0$ rejecting 
(in the following we often abbreviate $\ket{j}\in D$ by $j\in D$.) 
$$F':=\{\ket{i_2i_3\ldots i_kj};i_2,i_3,\ldots, i_k\in D, j\in F \}$$
is the set of accepting states of \CB'.
Let $V$ be a unitary transformation from $D'$ on itself 
that maps the vector $\ket{1\ldots 1}$ to the superposition (let $m:= w^{k-1}$):

\begin{multline}\llabel{VEq}
V \ket{1\ldots 1} =\\
=\frac{1}{\sqrt{2m}}\sum_{i_2,\ldots,i_k\in
  D}\ket{1i_2,\ldots,i_k}+
\frac{1}{2\sqrt{m}}\,\ket{t_0}\\+ \frac{\sqrt{2m-1}}{2\sqrt{m}}\,\ket{t_1}.
\end{multline}
The images of all other members of $D^\otimes$ are chosen such that
$V$ is unitary. 
This is possible, since $V \ket{1\ldots 1}$ is a vector of length $1$.

Let $i\in\{1,\ldots,n\}, \epsilon\in\{0,1\}$. 
The transformations 
$T_i^{\otimes,\epsilon}$ 
on $D^\otimes$ are defined according to Equation \ref{tcrossEq}. 
We define the unitary transformations $T_i^{'\epsilon}$ from $D'$ to
itself 
as behaving on $D^\otimes$ as  $T_i^{\otimes,\epsilon}$ and on $\{t_0,t_1\}$ as the identity:
\begin{math}
T_i^{'\epsilon}\ket{d}=T_i^{\epsilon}\ket{d}
\end{math}
for $d\in D^\otimes$,
\begin{math}
T_i^{'\epsilon}\ket{t_0}=\ket{t_0} 
\end{math}
and
\begin{math}
T_i^{'\epsilon}\ket{t_1}=\ket{t_1}.
\end{math}

Now \CB' is defined as computing according to 
$$
(V\cdot T_1^{'0}, V\cdot T_1^{'1}),(T_2^{'0},
T_2^{'1}),\ldots,(T_n^{'0}, T_n^{'1}).
$$

We determine the acceptance probability of the computation of 
$\CB'$ on an input $a$. \CB' starts with state $\ket{1\ldots 1}$. 
Applying $V$ on this start state has the result described in Equation \ref{VEq}.
Thus, applying 
$U'(a):=T^{a_{\sigma(n)}}_{n}\cdot \ldots\cdot
T^{a_{\sigma(2)}}_{2}\cdot 
T^{a_{\sigma(1)}}_{1}\cdot V$ 
on $\ket{1\ldots 1}$ has the result
\begin{multline*}
\frac{1}{\sqrt{2m}}\sum_{i_2,\ldots,i_k\in D}U'(a)\ket{1i_2,\ldots,i_k}\\
+\frac{1}{2\sqrt{m}}\,\ket{t_0}
+\frac{\sqrt{2m-1}}{2\sqrt{m}}\,\ket{t_1}.
\end{multline*}
The first part of this sum can be rewritten as 
\begin{multline*}
\frac{1}{\sqrt{2m}}\sum_{i_2,\ldots,i_k\in D}\sum_{j\in D}
\mu_{i_2,\ldots,i_kj}\ket{i_2,\ldots,i_kj}=\\
=\sum_{j\in D} \left( \frac{1}{\sqrt{2m}} \sum_{i_2,\ldots,i_k\in
    D}\mu_{i_2,\ldots,i_kj}\ket{i_2,\ldots,i_kj}\right)
\end{multline*} 
For the acceptance probability this implies 
\begin{multline*}
acc_{\CB'}(a)=\\
=\frac{1}{2m}\sum_{j\in F}\left| \sum_{i_2,\ldots,i_k\in D}
\mu_{i_2,\ldots,i_kj}\right|^2+\left|\frac{\sqrt{2m-1}}{2\sqrt{m}}\right|^2\\
=\frac{1}{2m}\sum_{j\in F}|\beta_j(a)|^2+\frac{2m-1}{4m},
\end{multline*}
where $\beta_j(a)$ is defined as in Equation \ref{accProbSumRepEQ}. 
Observe that the acceptance probability of the $k$-QOBDD \CB\, on $a$ is $\sum_{j\in F}|\beta_j(a)|^2$

Thus, if \CB\, accepts $a$ with probability at least $1/2$ then \CB'\, will do the same. We formulate this result as

\begin{theorem}\label{majOBDDTheorem}
For all natural numbers $k\geq 1$
a sequence of Boolean functions $(f_n)$ 
computable by polynomial size quantum $k$-OBDDs with unbounded error 
is also computable by  polynomial size quantum OBDDs. 
\end{theorem}

\bibliography{references}

\end{document}